\documentclass{article}
\usepackage{ijcai18}
\usepackage{helvet}  %Required
\usepackage{courier}  %Required
\usepackage{url}  %Required
\usepackage{graphicx}  %Required
\usepackage{amsmath}
\usepackage{amssymb}

\usepackage{bm}
\usepackage{xspace}

\usepackage{diagbox} 
\usepackage{cases}
\usepackage{enumerate}

\usepackage{times}
\usepackage{xcolor}
\usepackage{soul}

\usepackage{multicol}  
\usepackage{multirow} 
\usepackage{algorithm} 

\usepackage[utf8]{inputenc}
\usepackage[small]{caption}

\title{Interpretable Deep Learning Model for Online Multi-touch Attribution}

\author{
Dongdong Yang$^1$, 
Kevin Dyer$^2$, 
Senzhang Wang$^3$
\\ 
$^1$ Unversity of Southern California \\
$^2$ eBay Inc.  \\
$^3$ Nanjing University of Aeronautics and Astronautics\\
dongdony@usc.edu,
kkdyer8506@gmail.com,
szwang@nuaa.edu.cn
}

\begin{document}

\maketitle

\begin{abstract}
  In online advertising, users may be exposed to a range of different advertising campaigns, such as natural search or referral or organic search, before leading to a final transaction. Estimating the contribution of advertising campaigns on the user's journey is very meaningful and crucial. A marketer could observe each customer's interaction with different marketing channels and modify their investment strategies accordingly. Existing methods including both traditional last-clicking methods and recent data-driven approaches for the multi-touch attribution (MTA) problem lack enough interpretation on why the methods work. In this paper, we propose a novel model called DeepMTA, which combines deep learning model and additive feature explanation model for interpretable online multi-touch attribution. DeepMTA mainly contains two parts, the phased-LSTMs based conversion prediction model to catch different time intervals, and the additive feature attribution model combined with shaley values. Additive feature attribution is explanatory that contains a linear function of binary variables. As the first interpretable deep learning model for MTA, DeepMTA considers three important features in the customer journey: event sequence order, event frequency and time-decay effect of the event. Evaluation on a real dataset shows the proposed conversion prediction model achieves 91\% accuracy. %Meanwhile, DeepMTA can get channel importance for each customer journey instead of overall channel importance. This paper will open the gate of tackling MTA problem in two stages. The first stage is to build conversion prediction model and the other is to get touch attribution when interpreting the prediction model.
\end{abstract}

\section{Introduction}

In the digital world, ads reach everywhere and influence users across various channels such as paid search and social networks. Digital marketer observes each customer's interaction with different marketing channels. The advertisers could get users' feedback and modify their investment strategies accordingly. Attribution is critically important for online advertising so that the advertiser could know which channel attributes more. When advertisers know the contribution of their campaign touch points, they can make informed impression buying decisions based on the user’s final conversion. Aggregated attributions on advertising channels can provide useful guidance for advertisers to allocate budget on these advertising channels.

\begin{figure}
  \centering
  \includegraphics[width=.45\textwidth]{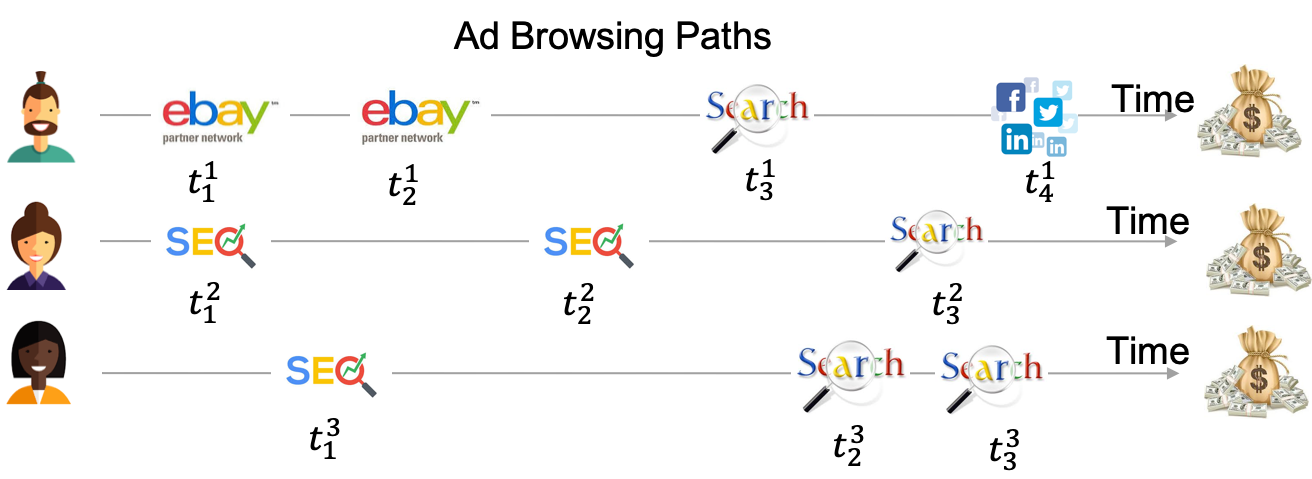}
  \caption{Customer journey illustrates three users' interactions with ad contents over typical channels.}
  \label{customer_journey_sample}
\end{figure}

Given the multi-touch events across a customer journey, the multi-touch attribution (MTA) model aims at analyzing the channels' exposure effect inside the customer impression journey on their purchase decision. Attribution model uses observations to estimate the exposure of marketing channels on conversion events. Capture exposure by modeling the impact of marketing channels on the likelihood of a conversion event. The exposure effects explored in \cite{shao2011data}, \cite{dalessandro2012causally} are captured by using the non-linear parametric models such as a logistic regression model. As shown in Figure \ref{customer_journey_sample}, users could reach a range of campaigns from multiple channels in their journeys before transaction. Thus it is important for advertisers to attribute the right conversion credit to each journey touch. The goal of MTA is to understand the importance of these touches. The touch across the customer journey could be regarded as an instance of a channel.

Data-driven methods have been explored in recent years, and have attracted rising research attention, unlike the tradtion rule-based methods such as last-click, linear and time decay model. For example, the most popular last-click model assumes that a user’s conversion is only caused by the last ad he clicked or viewed before. It gives a 100\% weight to the last clicking event and assigns all transaction value to it. To utilize the data for improving attribution models, Shao etc. \cite{shao2011data} proposed a data-driven multi-touch attribution model to allocate the contribution credit to different ad campaigns. \cite{abhishek2012media} proposed a hidden Markov model of an individual user's behavior based on conversion funnels to attribute conversion. \cite{dalessandro2012causally} utilized cooperative game theory with shapley value to approximately derive the channels attribution. \cite{ji2017additional} assumed that hazard rate which is defined in the time decaying across customer journey can be used to measure channel importance. \cite{ren2018learning} is a new neural network to tackle the MTA problem. %Different from the above models, our model has two stages. One is for building conversion prediction and the other is for the model to explicitly generate the touch attribution.
Deep learning models are also used to solve MTA problem in different ways recently \cite{ren2018learning}. \cite{ren2018learning} proposed dual-attention recurrent neural network to learn the attribution values through an attention mechanism directly from the conversion estimation objective. However, the major limitation of current deep learning models is that they cannot generate interpretable values. In order to meet the requirement of real applications, the MTA model should be both interpretable and catch enough information about the previous journal for further usage. %Considering the success of deep learning in recent years, we develop a deep learning model for MTA with interpretation. These previous data-driven models do not explore the problem in this way. There are two main reasons we consider about for nobody doing similar things: 1. the deep learning model is a black box before. It makes it hard to develop in marketing area which requires interpretation; 2. model itself should consider time decaying problem. 

\begin{figure}
  \centering
  \includegraphics[width=.45\textwidth]{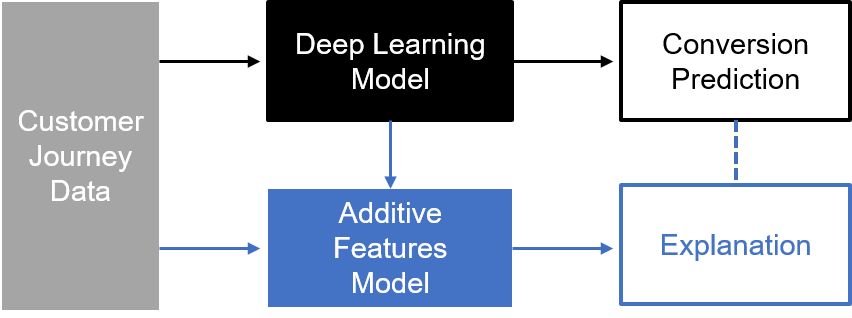}
  \caption{Architecture brief overview of DeepMTA.}
  \label{architecture-overview}
\end{figure}

\iffalse
\begin{table}[!htbp]
  \centering
  \begin{tabular}{|c|c|c|c|c|c|}
  \hline
  \diagbox{Model}{Features} & CF & CS & CDE & PP & CP \\ 
  \hline
  Last-clicking &&&&&\\
  \hline
  Shao et al.\cite{shao2011data} &&&&&\checkmark\\
  \hline
  Dalessandro et al \cite{dalessandro2012causally} &&&&\checkmark&\checkmark\\
  \hline
  E Anderl et al. \cite{anderl2016mapping} &\checkmark&\checkmark&&&\checkmark  \\
  \hline
  Abhishek et al. \cite{abhishek2012media} &\checkmark&\checkmark&&&\checkmark\\
  \hline
  N Barbieri et al. \cite{barbieri2016improving} &\checkmark&&\checkmark&&\checkmark\\
  \hline
  Ji et al. \cite{ji2017additional} &&\checkmark&\checkmark&&\checkmark\\
  \hline
  Deep MTA (ours) &\checkmark&\checkmark&\checkmark&&\checkmark  \\
  \hline
  \end{tabular}
  \caption{The table shows the features used by different models. CF means Channel Frequency; CS means Channel Sequence; CDE means Channel Decay Effect; PP means Prior Purchases; CP means Conversion Prediction. }
  \label{different_methods_features_differences}  

\end{table}
\fi

In this paper, we propose a novel method called DeepMTA, which combines deep learning model with the additive features explanation model. The framework of our proposal is shown in Figure \ref{architecture-overview}. The upper part is conversion prediction model which is implemented by phased-LSTMs. The bottom part is additive features explanation model with shaley values. The additive feature model is a linear function of a binary variable based on a power set, and then uses linear regression to calculate the importance of the features. The additional factor model is a simplified version of the Shapley value, which assigns a unique distribution of the total surplus generated by the coalition of all participants. The ultimate goal of these participants is to make customers buy goods. It is generally assumed that all players are independent. Additionally, DeepMTA considers three important features of customer journey: event sequence order, event frequency, time-decay effect. The code of this work is publicly available \footnote{\url{https://github.com/donzzzzy/interpretable-deep-model-mta}}.

\iffalse
The proposed model, DeepMTA, is the first deep learning framework to generate interpretation directly. The model we design is shown in the left part of Figure \ref{architecture}. The novel method uses the deep learning model and additive feature explanation model to generate the channel's importance with interpretation. We try to combine both deep learning and explanation model together to generate interpretable values for MTA. Our proposal mainly contains two stages. One is conversion prediction model using phased-LSTMs. Phased-LSTM is the new LSTM units that are able to catch different time intervals. The other part is to use additive features attribution models \cite{kim2017interpretability} combined with shaley values. Additive features attribution is an explanatory model that contains a linear function of binary variables. We combined it with shapley values in order to figure out the touch attribution. This paper will open the gate of tackling MTA problem in two stages. 
\fi

The major contributions of this work are summarized as follows:

\begin{itemize}
    \item We develop a novel interpretable deep learning model, DeepMTA, for online multi-touch attribution. To our knowledge, DeeMTA is the first model that combines deep learning and cooperative game theory.
    \item We fully consider three important features in the customer journey: event Sequence order, event frequency and time-decay effect, while previous works only consider parts of the features.
    \item We evaluate the conversion prediction model on a large real dataset with 91\% accuracy, and could calculate specific channels importance for each customer journey instead of whole weight for each channel.
\end{itemize}

The remainder of the paper is organized as follows. We first review related works in MTA, deep learning and model interpretation in Section 2. Then we present an overview of the model architecture and discuss the details of the model in Section 3 followed by the implementation details in Section 4. The experiments are conducted in Section 5. Finally, we conclude this work and discuss future work in Section 6.

%(1) developing a deep learning model with interpretation, using deep learning and cooperative game theory;

%(2) considering three important features in the customer journey: event Sequence order, event frequency, and time-decay effect;

%(3) evaluating the conversion prediction model with accuracy;

%(4) calculating specific channels importance for each customer journey.

\section{Related Work}

In online advertising, conversion attribution is usually calculated by rule-based methods. Recently, many research works for MTA have been proposed for modeling attribution of sequential touch points on various channels. \cite{shao2011data} proposed the first data-driven multi-touch attribution model based on a bagged Logistic regression model to estimate conversion rates based on ads viewed by users. The priority is to consider many customer journey features including channel frequency, channel sequence, channel attenuation effects and transition prediction. \cite{wooff2015time} considered the importance of time-weighted effects in the customer journey and designed a model for the problem of time decay. Based on extant marketing decision model acceptance literature, \cite{anderl2016mapping} put forward how to evaluate the model via channel credit allocation, accurate prediction of conversion events, stable and reproducible results, structure transparency and intuitive understanding, adaptability, and algorithm efficiency. \cite{berman2018beyond} proposed budget allocations for online campaigns by determining advertiser effectiveness and defining appropriate compensation methods.

\begin{figure*}
  \centering
  \includegraphics[width=1\textwidth]{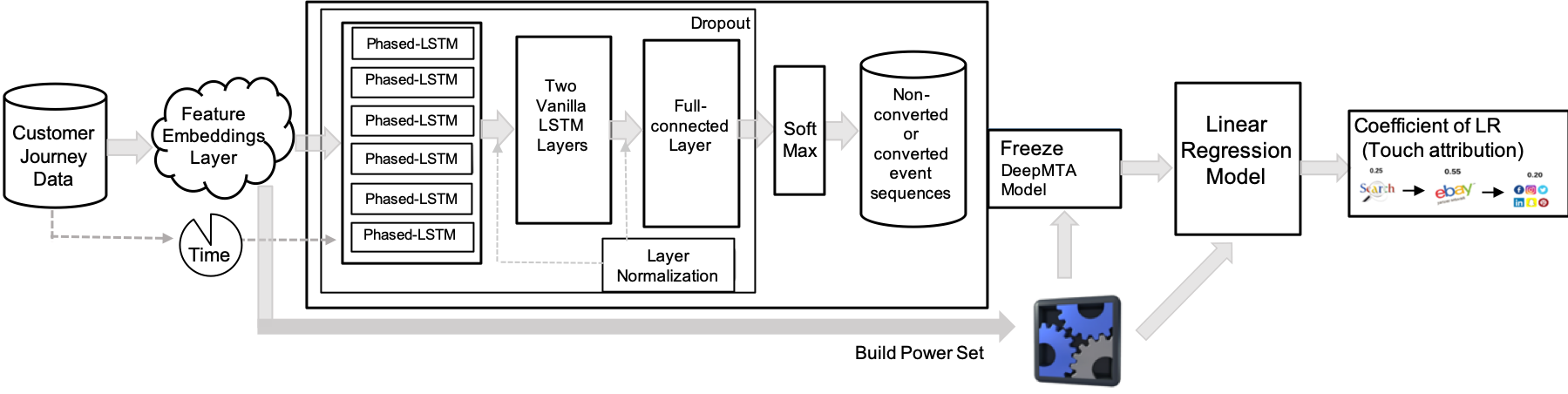}  
  \caption{Architecture details of DeepMTA.}  
  \label{architecture}  
\end{figure*}

Deep learning models have been widely applied in many areas such as NLP \cite{AAAI18}, data mining \cite{wang2019deep} and computer vision \cite{lecun2015deep}. All of the sequence deep learning models help to serialize customer streaming journey data. Compared with vanilla LSTMs, Phased-LSTM considers different time intervals and adds one more time gate in vanilla LSTM units. Thus it achieves faster convergence than vanilla LSTMs on tasks that need to learn long sequences. Furthermore, interpretation on deep neural networks is also developing fast, such as LIME \cite{ribeiro2016should} and SHAP \cite{lundberg2017unified}. They are easy to understand and could be applied to the features used by the model. The input variables of interpretable model may need to be different from the features. The LIME \cite{ribeiro2016should} method explains the predictions of each model based on a model approximation around a given prediction locally. The SHAP \cite{lundberg2017unified} method combines LIME and Shapely Values together to build a unified approach for model predictions. %Our intuition of interpretation of DeepMTA model is based on SHAP. 

\section{Framework}

In order to build an interpretable multi-touch attribution model in marketing, we use the linear regression as a bridge between deep learning black box and human understanding. Linear regression has been used in various scenarios for its easy understanding. The base of additive features explanation model is actually linear regression. 

The proposed DeepMTA contains two stages as shown in Figure \ref{architecture-overview}. The first stage is the deep learning model which aims to capture the information of the customer journey. The second stage is an additive features explanation model which is for interpreting the first stage part. It calculates the weight of each click touch based on shapley value \cite{shapley1953value} and linear regression. For more details, in the left part of Figure \ref{architecture}, deep learning model is built up with a kind of LSTM cells. In order to handle the clicking events with different timestamps, we use phased LSTMs \cite{neil2016phased} instead of regular LSTMs. 

After we train the deep learning model, the difficulty is to find the importance of sequence events of inputs. We need to know the importance of a clicking event and how much it contributes to the final prediction. It is to calculate the importance of the events instead of features inside the event. Briefly speaking, we build a powerset for additive feature attribution method, and then calculate the importance of the clicking events by using a linear regression. 

\subsection{Deep Learning Model}

Given a customer journey, predicting the prediction possibility of conversion. Since the customer journey is built up with a sequence of clicking events, we intuitively model the scenarios by using RNN to consider channel frequency - the number of repetitive occurrences, channel sequence - order of occurrence and channel decay effect - channel exposure influence into conversion. To learn the customer journey information, we regard this problem as the prediction problem for conversion.

For feature embedding part of deep model, the features include channel id, campaign id and timestamp interval. The considered attributes should have relationship to click event itself. When interpreting the customer journey, some click event will be removed away from the journey for measuring the effect of it. Imaging that a removed event is critical, it means the prediction accuracy will drop rapidly. Otherwise the prediction will not be affected much. Therefore, if we set the features that are not related to the event itself, the final effect will also contain the effect of those unrelated features on the prediction.

The deep model considers the sequence of events across the customer journey as the input. One channel, eg. paid search, could occur many times and the same channel could be followed by the same channel. \cite{anderl2014mapping} aggregate many same channels into one for their customer journey when they build the Markov graph based on their assumptions. To consider the time decaying effect and not to aggregate the same channels, we use phased-lstm cells \cite{neil2016phased}, instead of vanilla LSTM, to learn the sequence of the customer journey. To catch the different time intervals, we use phased LSTMs with a additional time gate, compared with vanilla LSTMs. Time decaying problem means most recent events should have more impacts on the final result. \cite{ji2017additional} used the corresponding hazard rate to reflect the influence of an ad exposure for the time decaying problem. But we want RNN cells itself could handle time decaying issue. Then the model is evaluated by conversion prediction. 

\begin{align}
  \label{eq: phased_calulation1}
  i_t &= \sigma_i(x_tW_{xi} + h_{t-1}W_{hi} + w_{ci} \odot c_{t-1} + b_i ) \\ 
  f_t &= \sigma_f(x_tW_{xf} + h_{t-1}W_{hf} + w_{cf} \odot c_{t-1} + b_f ) \\
  \widetilde{c_t} &= f_t \odot c_{t-1} + i_t \odot \sigma_c(x_tW_{xc} + h_{t-1}W_{hc} + b_c ) \\
  c_t &= k_t \odot \widetilde{c_t} + (1-k_t) * c_{t-1} \\
  o_t &= \sigma_o(x_tW_{xo} + h_{t-1}W_{ho} + w_{co} \odot c_{t-1} + b_f ) \\
  \widetilde{h_t} &= o_t \odot \sigma_h(\widetilde{c_t}) \\
  h_t &= k_t \odot \widetilde{h_t} + (1-k_t) * h_{t-1}
  \label{eq: phased_calulation7} 
\end{align}

The Phased LSTM cell extends the LSTM cell by adding a new time gate. The calculation is shown in Equation \ref{eq: phased_calulation1} - \ref{eq: phased_calulation7}. Three parameters control the opening and closing of the door. The cell state $c_t$ and $h_t$ are triggered only when the gate is open. $\tau$ controls the period of the oscillation. $r_{on}$ controls the ratio of the duration of the open phase. $s$ controls the phase shift of the oscillation to each LSTM cell. The gates use $\sigma_i$, $\sigma_f$, $\sigma_o$ and $\sigma_c$, $\sigma_h$. The cell state $c_t$ itself is updated by the previous cell state. The previous cell state is controlled by $f_t$. Input state $\odot$ is from the element-wise product of $i_t$. All parameters can be learned during the training process.

The gate $k_t$ has three phases as shown in Equation \ref{eq: phased_calulation8} - \ref{eq: phased_calulation11}, where $\tau$ is the period, $s$ is the phase shift, and $r_{on}$ is the ratio of the open period to the total period $\tau$. In the first two stages, the gate openness increases from 0 to 1, and then decreases from 1 to 0. In the third phase, the gate is closed and the previous battery state is maintained. Leaks with a rate are active during the closing phase and propagate important gradient information even when the gate is closed. During the opening phase of the time gate, the linear slope of $k_t$ allows the error gradient to be transmitted efficiently.

\begin{equation}
  \phi_t = \frac{(t-s) mod \tau}{\tau}
  \label{eq: phased_calulation8} 
\end{equation}

\begin{numcases}{k_t=}
  \frac{2\phi_t}{r_{on}}, &$\phi_t < \frac{1}{2}r_{on}$\\
  2-\frac{2\phi_t}{r_{on}}, &$\frac{1}{2}r_{on} < \phi_t < r_{on}$ \\
  \alpha \phi_t, &otherwise
  \label{eq: phased_calulation11} 
\end{numcases}

\iffalse
\begin{figure}
  \centering
  \includegraphics[width=.40\textwidth]{phased-lstm.png}  
  \caption{Phased-LSTM cell in our model.}  
  \label{fig: phased}  
\end{figure}
\fi

The objective function of our model uses softmax and cross-entropy function as shown in Equation \ref{eq: goal_function}, where $M$ is the number of clicking events in a customer journey, $y_i'$ is the ground-truth label, and $y_i$ is the predicted label value after softmax function. For each event across the customer journey, the output will be 0 or 1. 0 means non-conversion and 1 means conversion. After the model converges, we freeze the model for the usage in the interpretation part.

\begin{equation}
  H_{y'}(y) = - \sum_i^M (y_i'log(y_i) + (1-y_i')log(1-y_i))
  \label{eq: goal_function} 
\end{equation}

\subsection{Interpretation}

To understand interpretation model part, we need to distinguish interpretable data representations from event characteristics. Interpretable model requires the use of human-understandable representations, regardless of the actual functions used by the model. Taking the classification problem as an example, the possible interpretable data representation is a binary vector. Even though the classifier may use more complex different features inside the event, the binary vector indicates the presence or absence of the event. We denote $x \in {R}^d$ as the original variable which needs to be explained, and $x' \in \{0, 1\}$ to be a binary variable for the interpretation.

In this part, the goal is to calculate the importance of different events across the customer journey with interpretation. After we train the conversion prediction model as shown in the left part of Figure \ref{architecture}, we use additive features explanation model to get the importance of events as shown in the right of the Figure \ref{architecture}. Three previous methods use classic equations from shapley value to compute explanations of model predictions: shapley regression values \cite{lipovetsky2001analysis}, shapley sampling values \cite{anderl2014mapping} and quantitative input influence \cite{datta2016algorithmic}. In our method, we use the shapley regression value for shorter customer journeys and the shaley sampled value for longer journeys. Shapley value method is a general credit distribution method in cooperative game theory. It is based on assessing the marginal contribution of each player in the game. Points are assigned to each individual player. The rough value is the expected value of the marginal contribution over all possible permutations of the player.

\begin{equation}
  \label{eq:shaply_value}
  \phi_i = \sum_{S \subseteq F \setminus \{i\}} \frac{|S|!(|F| - |S| - 1)!}{|F|!} [f_{S \cup \{i\}}(x_{S \cup \{i\} }) - f_S(x_S)]
\end{equation}

Shapley regression values, in Equation \ref{eq:shaply_value}, represent multicollinearity, the characteristic importance of linear models. This method requires retrain the model on all feature subsets $S \subseteq F$, where $F$ is the set of all features. It assigns an weight to each feature. The weight of feature represents the impact on model predictions. To compute this effect, a model $f_{S \cup \{i\}}$ is trained with that feature presentation, and another model $f_S$ is trained with the feature withheld. Then, predictions from the two models are compared on the current input $f_{S \cup \{i\}} (x_{S \cup \{i\}}) - f_S(x_S)$, where $x_S$ represents the values of the input features in the set $S$. Because the effect of retaining features depends on other features in the model, previous differences are calculated for all possible subsets $S \subseteq F \setminus \{i\} $. The Shapley value is then calculated and used as a feature attribution. They are a weighted average of all possible differences. For Shapley regression values, $h_x$ will map 1 or 0 to the original input space, where 1 means the input is included in the model and 0 means the exclusion from the model. If we let $\phi_0 =f_{\phi}(\phi)$, the Shapley regression values match the additive feature attribution method, as Equation \ref{eq:additive} shows. In \ref{eq:additive}, $z'\in \{0, 1\}^M$, $M$ denotes the number of clicking events in a customer journey, and $\phi_i \in {R}$, {R} denotes real number set. 

\begin{equation}
  \label{eq:additive}
  g(z') = \phi_0 + \sum_{i=1}^M \phi_i z_i'
\end{equation}

The additive feature attribution method has an explanatory model, which is a linear function of binary variables. An weight $\phi_i$ is for each feature, and the effects of all feature attributions are added to get the output of the original model.

  \begin{equation}
    Y_{acc} = \sum_i^M (I(y_i', y_i) \odot mask^T)
    \label{eq: acc_function} 
  \end{equation}

The interpretation part of the model contains the following three steps. First, get the powerset of input and generate the mask matrix $X_{mask}$. The size of the mask matrix is [$2^n$, $n$], where $n$ is the number of events in a customer engagement journey. This part is to mask the events across the customer journey to evaluate their effects. If the event is masked, the corresponding position in the mask matrix is set to 0; otherwise, the corresponding position is set to 1. The input of the conversion prediction model is $X$=[$x_1$, $x_2$, $x_3$, ..., $x_n$], where $x_n$ is 1 or 0, denoting a customer journey containing a sequence of clicking events. Second, generate the new input using the mask matrix from the first step, put it into the conversion prediction model generated to get the prediction $Pred$. Then by comparing it to the ground truth label $Label$, we could know the effect of missing events. Finally, obtain the accuracy $Y_{acc}$ for a specific input. In details, one row of the mask matrix $X_{mask}$ is $mask = [1, 1, 1, ..., 1, 0]^T$, where the size of the $mask$ is [1, $n$]. In this case, only the final event is masked and set to empty. Then the real input is $X_{real}= X \odot mask$. After putting it as an input of conversion model, $Pred$ is obtained, where the size of $Pred$ is [$n$, 1]. When we do a comparison between $Pred$ and $Label$, the last digit of output is ignored since the last event of input is ignored. In this way, we could get $Y_{acc}$. The calculation equation is given in Equation \ref{eq: acc_function}, where $I$ is an indicator function, $y_i'$ is the ground-truth label and $y_i$ is the predicted label value after softmax function. Finally, Use linear regression to calculate the weight. $X$$W$ = $Y_{acc}$. $Y_{acc}$ is the prediction accuracy and $X$ is the mask matrix. For simplicity, we assume that every clicking event across the customer journey is independent to each other.

\iffalse
\begin{enumerate}[step 1]
  \item 
  \item 
  \item 
\end{enumerate}
\fi

\section{Implementation Details}

\subsection{Sampling}
Sampling is to generated data and randomly select a fixed number of samples. The Shapley sample values \cite{anderl2014mapping} aims to explain any model by (1) applying the sample approximation to the equation \ref{eq:shaply_value}, and (2) approximating the estimates from the model by removal effect of a variable. This eliminates the need to retrain the model and allows calculation of differences less than $2^{|F|}$.

\subsection{Model training }
To avoid model over-fitting, we apply dropout and layer normalization. Dropout \cite{srivastava2014dropout} is the random removal of units or their connections from the neural network during training. This prevents too much mutual adaptation between the units. The layer normalization method is to calculate the normalized statistics at each time step. The inputs in the recurrent layer are calculated from the current input $x_t$ and previous vector of hidden states $h_{t-1}$ which are calculated as $\alpha_t = W_{hh} h_{t-1} + W_{xh} x_t$. 

\iffalse
Layer-normalized recursive layer re-centers and re-scales its activation in Equation \ref{eq: layer_normalization}, where $u_t = \frac{1}{H} \sum_{i=1}^H a_t^i $, $\sigma_t = \sqrt{\frac{1}{H} \sum_{i=1}^H (a_t^i-\mu_t)^2}$

\begin{equation}
  h_t = f[\frac{g}{\sigma_t} \odot (\alpha_t - \mu_t) + b]
  \label{eq: layer_normalization} 
\end{equation}
\fi

\subsection{Weight Clipping}
The coefficent of linear regression might be negative, a filter is needed to ignore the negative weights before normalizing all weights. The sum of weights is equal to 1 through $W_{new} = Norm(filter(W))$, where $W$ is the matrix we calculate for each customer journey, $filter$ is a function to ignore negative values and $Norm$ is a normalization function.

\begin{center}
\begin{tabular}{ll}
\hline Paramters & Values \\
\hline
Class Number & 2 \\
Sequence Length & 32 \\
Dropout & 0.5 \\
Hidden Layer & 2 \\
Hidden Cell Units & 1024 \\
Batch Size & 128 \\
Learning Rate & 0.01 \\
Epoch Number & 300 \\
\hline
\label{table:hyperparamters}
\end{tabular}
\end{center}

\section{Experiment}
We conduct our experiments on a real large customer journey dataset. The evaluation metrics is the accuracy of conversion prediction in the ROC graph and comparing total GMV result for each channel with the last-clicking method.

\subsection{Dataset}

\iffalse
\begin{figure}
  \centering
  \includegraphics[width=.45\textwidth]{channel_trasmit.png}  
  \caption{How our customers transmit from channel to channel in our dataset.}  
  \label{fg:channel_trasmit}  
\end{figure}
\fi

The data is collected on April 25th, 2018 from eBay dataware house. We trace back 10 days for each customer and record all events. Based on the transaction, a streaming customer journey is split into several customer journeys. The original dataset for each customer is a steaming of click events. Then based on conversion event, these events are split into separate journeys. In our experiments, the dataset only contains conversion journey without non-conversion in order to make the dataset representative. After analysis, the conversion events are less than 1\%. 

In our experiments, 100k customer journeys are generated, 90\% of which are used for training and validation and the remaining 10\% are used for testing. For 90k training customer journeys, it contains 742,201 clicking events; for 10k testing customer journeys, it contains 81,530 clicking events. For each event, the attributes contains channel id (paid social, paid search, SEO, affiliates, etc.), user id, transaction timestamps, transaction type (mobile, desktop), click URL, publisher id, landing page URL, click event rank and transaction id. To build the customer journey, we group the events by user id and sort them by time. Finally based on the clicking event rankings, the clicking streaming is split into multiple journeys.

\subsection{Parameters Setting}

Table \ref{table:hyperparamters} shows the parameter settings. We set some parameters empirically, such as the batch size, dropout rate, the number of epochs. We set the learning rate as $10^{-2}$, which is chosen from \{$10^{-1}$, $10^{-2}$, $10^{-3}$, $10^{-4}$, $10^{-5}$\}. We select 1024 LSTM units from the set \{256, 512, 1024, 2048\}. Training epoch is set to 300 for model converge. The class number is 2 since for each step of a sequence it only generates two states. The batch size is set to 128 which is selected from \{32, 64, 128, 256, 512, 1024\} so that the model does not converge too fast or too slow.

\begin{figure}
  \centering
  \includegraphics[width=.45\textwidth]{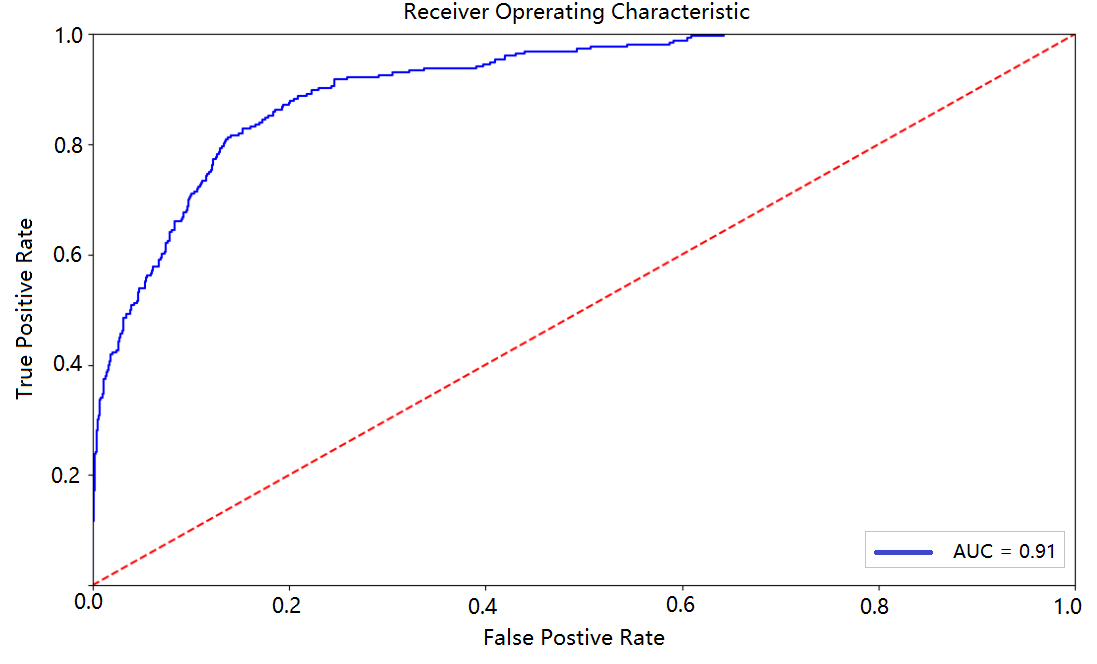}  
  \caption{ROC graph of our method.}  
  \label{ROC_Curve}  
\end{figure}

\subsection{Experiment Results and Analysis}

\begin{figure}
  \centering
  \includegraphics[width=.45\textwidth]{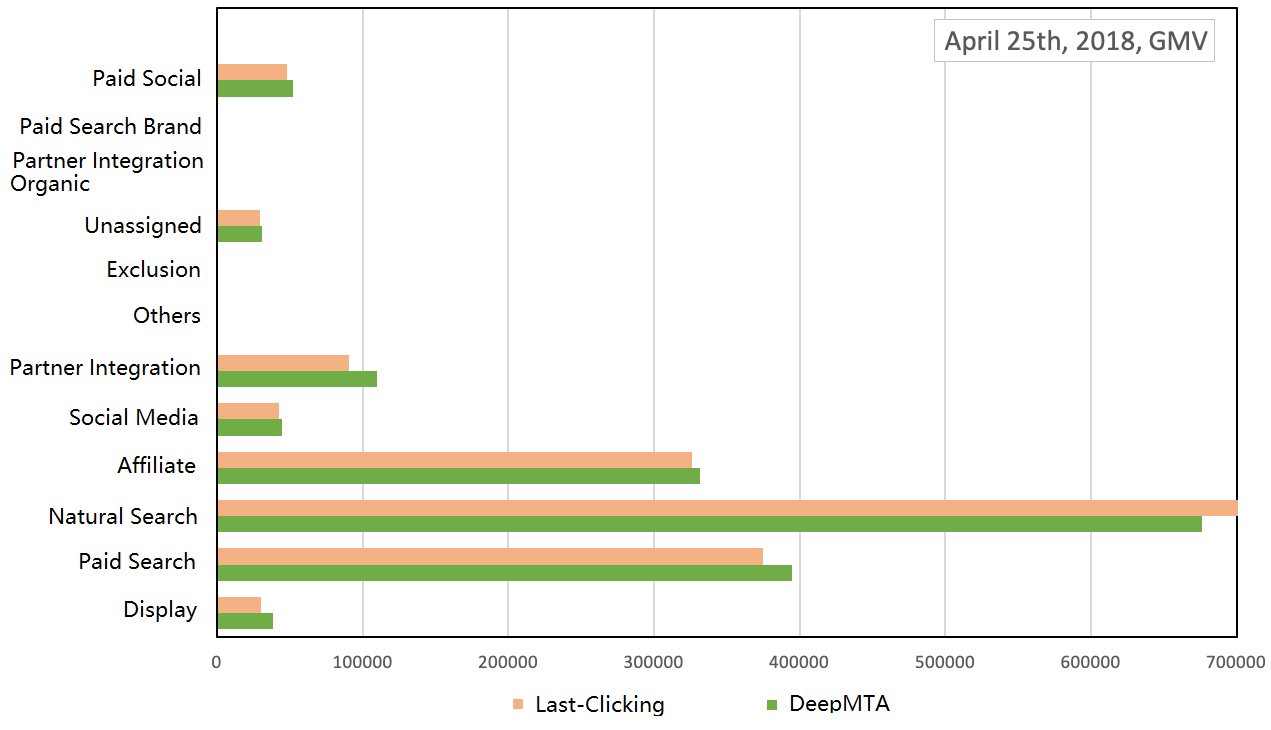}  
  \caption{GMV comparison between the proposed DeepMTA and the last-clicking attribution method.}  
  \label{fg:Comparison}  
\end{figure}

In this part, we get an overview of DeepMTA performance. Following the previous work \cite{shao2011data}, which proposed the data-driven method in MTA and 6 criteria evaluation methods in related work, we evaluate our method in two ways. The first is to get the ROC to evaluate the model conversion prediction performance, and the other one is to gather the Gross Merchandise Value (GMV) and then compare our results with the results of last clicking method. GMV is a value number used in online retailing and represents the total sales value of goods sold through a specific market within a period.

The ROC curve of our model is showed in Figure \ref{fg:Comparison}. One can see that the ROC of our model reaches AUC 91\%, which is a rather good performance. Particularly, when the false position  reaches 0.65, the true positive is 1.00. It shows the tradeoff between sensitivity and specificity (any increase in sensitivity will be accompanied by a decrease in specificity). The closer the curve is to the left and upper boundaries of the ROC space, the more accurate the test. The closer the curve is to the 45-degree diagonal of the ROC space, the lower the accuracy of the test.

Second, following most previous works in MTA that compare their results with last-clicking method, w{}e also compare our result with the last-clicking model since it is the most popular model in the industry. Many other models are not guaranteed to work very well and most of the model using deep learning directly is not interpretable. After we calculate the weight for each channel and manipulate on the raw weight followed by normalization, we are able to allocate the GMV of each customer journey to each channel.

The total average accumulative attribution for each channel are as followed: Display is 0.017, Paid Search is 0.205, Natural Search is 0.321, Affiliate is 0.143, Social Media is 0.018, Partner Integration is 0.071, Paid Social is 0.034 and Other are 0.014. Note that the sum of the weights of all channels is not equal to 1. After we calculate channel importance for each customer journey, if the weight of that channel in a customer journey is negative, we just set it 0.0 and then normalize all other positive values. We set these normalized values as the actual channel importance in that customer journey. Then we could get a total weight for a channel. We sum up all channels importance of all customer journey and average it based on the number of customer journeys that this channel occurs to get the average accumulative attribution. We set these normalized values to the channel and allocate the GMV to this channel finally. After normalizing all calculated weights, we then use them to allocate the GMV value. 

\iffalse
\begin{center}

  \begin{tabular}{ll}
  \hline Channel & Aggregate Weight \\
  \hline
  Display & 0.017 \\
  Paid Search & 0.205 \\
  Natural Search & 0.321 \\
  Affiliate & 0.143 \\
  Social Media & 0.018 \\
  Partner Integration & 0.071 \\
  Partner Integration - Organic & 0.0 \\
  Paid Search - Brand & 0.0 \\
  Unassigned & 0.014 \\
  Exclusion & 0.0 \\
  Paid Social & 0.034 \\
  Others & 0.0 \\
  \hline
  \label{table:channelweight}
  \end{tabular}
  \end{center}

\begin{figure}
  \centering
  \includegraphics[width=.45\textwidth]{weight_per_channel.png}  
  \caption{Channels Weight of our method.}  
  \label{weight_per_channel}  
\end{figure}
\fi

\subsection{Case Analysis}

The first step contains data prepossessing and model training. Assuming the customer journey is $A_1 \rightarrow B_1 \rightarrow A_2 \rightarrow A_3 \rightarrow B_2 $, where $A_i$ is a list of paid search click events, $B_i$ is a list of natural search click events. Based on the features of company id, channel id and time stamp, we embed $A$ as $x_A$, $B$ as $x_B$, where each $x_i = [v_{company_i}, v_{channel_i}, t_i]$, $v_{company}$ and $v_{channel}$ is a one-hot embedding vector, $t_i$ is the time period shifting from $t_0$. The label of this customer journey $label$ is $[0, 0, 0, 0, 1]$, where 0 means non-converted, 1 means converted. Then based on these three features of the samples and their labels, we train a conversion prediction model. 

\begin{figure}
  \centering
  \includegraphics[width=.45\textwidth]{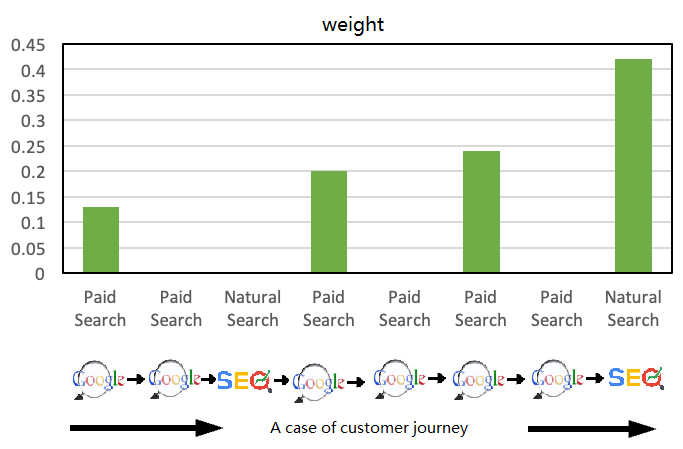}  
  \caption{A customer journey case. The negative weights have been removed and the positive weights have been normalized.}  
  \label{fig:wight_case1}  
\end{figure}

The next step is to know how important of the input. Mask matrix is generated to feed masked input to the deepMTA for training. In this case of $A_1 \rightarrow B_1 \rightarrow A_2 \rightarrow A_3 \rightarrow B_2 $, mask matrix powerset is built first. Then a sample for mask matrix is like $maskMatrix_{8} = [0, 0, 1, 1, 1]$. The generated sample $x_{new}$ will be $x$ * $maskMatrix_{8}^T$. At the same time, the label $label_{new}$ will be $label$ * $maskMatrix_{8}$. Then we put $x_new$ into conversion prediction model and get a predicted labels $y_{pred}$. After multiplied by $maskMatrix_{8}$, it is compared with $y_{new}$. After getting accuracy for this sample, linear regression could be used to calculate the importance of input clicking events. $maskMatrix W = accMatrix$, where $W$ is the importance of clicking events. The final step is to normalize the weights.

\iffalse
\begin{figure}
  \centering
  \includegraphics[width=.45\textwidth]{wight_case2.png}  
  \caption{2nd over 4 customer journey examples.}  
  \label{wight_case2}  
\end{figure}

\begin{figure}
  \centering
  \includegraphics[width=.45\textwidth]{wight_case3.png}  
  \caption{3rd over 4 customer journey examples.}  
  \label{wight_case3}  
\end{figure}

\begin{figure}
  \centering
  \includegraphics[width=.45\textwidth]{wight_case4.png}  
  \caption{4th over 4 customer journey examples.}  
  \label{wight_case4}  
\end{figure}
\fi

\section{Conclusion}
We built up a unified architecture of deep learning for multi-touch attribution with interpretation.  It is the first trial in this area. For the consideration of different time intervals, phased-LSTM is used so that the model could catch different time intervals information. After the model is trained well, which means DeepMTA achieves a high accuracy in ROC in conversion prediction, an additive explaination model is involved to generate the multi-touch attributions. These generated attributions for each channel could be well interpreted in a mathematical way, which makes it interpretable.

\iffalse
%\begin{comment}
\section{Acknowledgments}
Thanks.
%\end{comment}
\fi

\bibliographystyle{named}
\bibliography{ijcai18}

\begin{thebibliography}{}

\bibitem[\protect\citeauthoryear{Abhishek \bgroup \em et al.\egroup
  }{2012}]{abhishek2012media}
Vibhanshu Abhishek, Peter Fader, and Kartik Hosanagar.
\newblock Media exposure through the funnel: A model of multi-stage
  attribution.
\newblock 2012.

\bibitem[\protect\citeauthoryear{Anderl \bgroup \em et al.\egroup
  }{2014}]{anderl2014mapping}
Eva Anderl, Ingo Becker, Florian~V Wangenheim, and Jan~Hendrik Schumann.
\newblock Mapping the customer journey: A graph-based framework for online
  attribution modeling.
\newblock 2014.

\bibitem[\protect\citeauthoryear{Anderl \bgroup \em et al.\egroup
  }{2016}]{anderl2016mapping}
Eva Anderl, Ingo Becker, Florian Von~Wangenheim, and Jan~Hendrik Schumann.
\newblock Mapping the customer journey: Lessons learned from graph-based online
  attribution modeling.
\newblock {\em International Journal of Research in Marketing}, 33(3):457--474,
  2016.

\bibitem[\protect\citeauthoryear{Barbieri \bgroup \em et al.\egroup
  }{2016}]{barbieri2016improving}
Nicola Barbieri, Fabrizio Silvestri, and Mounia Lalmas.
\newblock Improving post-click user engagement on native ads via survival
  analysis.
\newblock In {\em Proceedings of the 25th International Conference on World
  Wide Web}, pages 761--770. International World Wide Web Conferences Steering
  Committee, 2016.

\bibitem[\protect\citeauthoryear{Berger \bgroup \em et al.\egroup
  }{2002}]{berger2002marketing}
Paul~D Berger, Ruth~N Bolton, Douglas Bowman, Elten Briggs, Vasanth Kumar, Arun
  Parasuraman, and Creed Terry.
\newblock Marketing actions and the value of customer assets: A framework for
  customer asset management.
\newblock {\em Journal of Service Research}, 5(1):39--54, 2002.

\bibitem[\protect\citeauthoryear{Berman}{2018}]{berman2018beyond}
Ron Berman.
\newblock Beyond the last touch: Attribution in online advertising.
\newblock {\em Marketing Science}, 37(5):771--792, 2018.

\bibitem[\protect\citeauthoryear{Cho \bgroup \em et al.\egroup
  }{2014}]{cho2014learning}
Kyunghyun Cho, Bart Van~Merri{\"e}nboer, Caglar Gulcehre, Dzmitry Bahdanau,
  Fethi Bougares, Holger Schwenk, and Yoshua Bengio.
\newblock Learning phrase representations using rnn encoder-decoder for
  statistical machine translation.
\newblock {\em arXiv preprint arXiv:1406.1078}, 2014.

\bibitem[\protect\citeauthoryear{Chung \bgroup \em et al.\egroup
  }{2014}]{chung2014empirical}
Junyoung Chung, Caglar Gulcehre, KyungHyun Cho, and Yoshua Bengio.
\newblock Empirical evaluation of gated recurrent neural networks on sequence
  modeling.
\newblock {\em arXiv preprint arXiv:1412.3555}, 2014.

\bibitem[\protect\citeauthoryear{Collobert and
  Weston}{2008}]{collobert2008unified}
Ronan Collobert and Jason Weston.
\newblock A unified architecture for natural language processing: Deep neural
  networks with multitask learning.
\newblock In {\em Proceedings of the 25th international conference on Machine
  learning}, pages 160--167. ACM, 2008.

\bibitem[\protect\citeauthoryear{Dalessandro \bgroup \em et al.\egroup
  }{2012}]{dalessandro2012causally}
Brian Dalessandro, Claudia Perlich, Ori Stitelman, and Foster Provost.
\newblock Causally motivated attribution for online advertising.
\newblock In {\em Proceedings of the Sixth International Workshop on Data
  Mining for Online Advertising and Internet Economy}, page~7. ACM, 2012.

\bibitem[\protect\citeauthoryear{Datta \bgroup \em et al.\egroup
  }{2016}]{datta2016algorithmic}
Anupam Datta, Shayak Sen, and Yair Zick.
\newblock Algorithmic transparency via quantitative input influence: Theory and
  experiments with learning systems.
\newblock In {\em Security and Privacy (SP), 2016 IEEE Symposium on}, pages
  598--617. IEEE, 2016.

\bibitem[\protect\citeauthoryear{Gers \bgroup \em et al.\egroup
  }{1999}]{gers1999learning}
Felix~A Gers, J{\"u}rgen Schmidhuber, and Fred Cummins.
\newblock Learning to forget: Continual prediction with lstm.
\newblock 1999.

\bibitem[\protect\citeauthoryear{Geyik \bgroup \em et al.\egroup
  }{2014}]{geyik2014multi}
Sahin~Cem Geyik, Abhishek Saxena, and Ali Dasdan.
\newblock Multi-touch attribution based budget allocation in online
  advertising.
\newblock In {\em Proceedings of the Eighth International Workshop on Data
  Mining for Online Advertising}, pages 1--9. ACM, 2014.

\bibitem[\protect\citeauthoryear{Ji and Wang}{2017}]{ji2017additional}
Wendi Ji and Xiaoling Wang.
\newblock Additional multi-touch attribution for online advertising.
\newblock In {\em AAAI}, pages 1360--1366, 2017.

\bibitem[\protect\citeauthoryear{Ji \bgroup \em et al.\egroup
  }{2016}]{ji2016probabilistic}
Wendi Ji, Xiaoling Wang, and Dell Zhang.
\newblock A probabilistic multi-touch attribution model for online advertising.
\newblock In {\em Proceedings of the 25th ACM International on Conference on
  Information and Knowledge Management}, pages 1373--1382. ACM, 2016.

\bibitem[\protect\citeauthoryear{Jordan \bgroup \em et al.\egroup
  }{2011}]{jordan2011multiple}
Patrick Jordan, Mohammad Mahdian, Sergei Vassilvitskii, and Erik Vee.
\newblock The multiple attribution problem in pay-per-conversion advertising.
\newblock In {\em International Symposium on Algorithmic Game Theory}, pages
  31--43. Springer, 2011.

\bibitem[\protect\citeauthoryear{Kingma and Ba}{2014}]{kingma2014adam}
Diederik~P Kingma and Jimmy Ba.
\newblock Adam: A method for stochastic optimization.
\newblock {\em arXiv preprint arXiv:1412.6980}, 2014.

\bibitem[\protect\citeauthoryear{Kira and Rendell}{1992}]{kira1992practical}
Kenji Kira and Larry~A Rendell.
\newblock A practical approach to feature selection.
\newblock In {\em Machine Learning Proceedings 1992}, pages 249--256. Elsevier,
  1992.

\bibitem[\protect\citeauthoryear{LeCun \bgroup \em et al.\egroup
  }{2015}]{lecun2015deep}
Yann LeCun, Yoshua Bengio, and Geoffrey Hinton.
\newblock Deep learning.
\newblock {\em nature}, 521(7553):436, 2015.

\bibitem[\protect\citeauthoryear{Lee \bgroup \em et al.\egroup
  }{2012}]{lee2012estimating}
Kuang-chih Lee, Burkay Orten, Ali Dasdan, and Wentong Li.
\newblock Estimating conversion rate in display advertising from past
  erformance data.
\newblock In {\em Proceedings of the 18th ACM SIGKDD international conference
  on Knowledge discovery and data mining}, pages 768--776. ACM, 2012.

\bibitem[\protect\citeauthoryear{Li and Kannan}{2014}]{li2014attributing}
Hongshuang Li and PK~Kannan.
\newblock Attributing conversions in a multichannel online marketing
  environment: An empirical model and a field experiment.
\newblock {\em Journal of Marketing Research}, 51(1):40--56, 2014.

\bibitem[\protect\citeauthoryear{Lipovetsky and
  Conklin}{2001}]{lipovetsky2001analysis}
Stan Lipovetsky and Michael Conklin.
\newblock Analysis of regression in game theory approach.
\newblock {\em Applied Stochastic Models in Business and Industry},
  17(4):319--330, 2001.

\bibitem[\protect\citeauthoryear{Lundberg and Lee}{2017}]{lundberg2017unified}
Scott~M Lundberg and Su-In Lee.
\newblock A unified approach to interpreting model predictions.
\newblock In {\em Advances in Neural Information Processing Systems}, pages
  4765--4774, 2017.

\bibitem[\protect\citeauthoryear{Neil \bgroup \em et al.\egroup
  }{2016}]{neil2016phased}
Daniel Neil, Michael Pfeiffer, and Shih-Chii Liu.
\newblock Phased lstm: Accelerating recurrent network training for long or
  event-based sequences.
\newblock In {\em Advances in Neural Information Processing Systems}, pages
  3882--3890, 2016.

\bibitem[\protect\citeauthoryear{Ren \bgroup \em et al.\egroup
  }{2018}]{ren2018learning}
Kan Ren, Yuchen Fang, Weinan Zhang, Shuhao Liu, Jiajun Li, Ya~Zhang, Yong Yu,
  and Jun Wang.
\newblock Learning multi-touch conversion attribution with dual-attention
  mechanisms for online advertising.
\newblock In {\em Proceedings of the 27th ACM International Conference on
  Information and Knowledge Management}, pages 1433--1442. ACM, 2018.

\bibitem[\protect\citeauthoryear{Ribeiro \bgroup \em et al.\egroup
  }{2016}]{ribeiro2016should}
Marco~Tulio Ribeiro, Sameer Singh, and Carlos Guestrin.
\newblock Why should i trust you?: Explaining the predictions of any
  classifier.
\newblock In {\em Proceedings of the 22nd ACM SIGKDD international conference
  on knowledge discovery and data mining}, pages 1135--1144. ACM, 2016.

\bibitem[\protect\citeauthoryear{Richardson \bgroup \em et al.\egroup
  }{2007}]{richardson2007predicting}
Matthew Richardson, Ewa Dominowska, and Robert Ragno.
\newblock Predicting clicks: estimating the click-through rate for new ads.
\newblock In {\em Proceedings of the 16th international conference on World
  Wide Web}, pages 521--530. ACM, 2007.

\bibitem[\protect\citeauthoryear{Shao and Li}{2011}]{shao2011data}
Xuhui Shao and Lexin Li.
\newblock Data-driven multi-touch attribution models.
\newblock In {\em Proceedings of the 17th ACM SIGKDD international conference
  on Knowledge discovery and data mining}, pages 258--264. ACM, 2011.

\bibitem[\protect\citeauthoryear{Shapley}{1953}]{shapley1953value}
Lloyd~S Shapley.
\newblock A value for n-person games.
\newblock {\em Contributions to the Theory of Games}, 2(28):307--317, 1953.

\bibitem[\protect\citeauthoryear{Srivastava \bgroup \em et al.\egroup
  }{2014}]{srivastava2014dropout}
Nitish Srivastava, Geoffrey Hinton, Alex Krizhevsky, Ilya Sutskever, and Ruslan
  Salakhutdinov.
\newblock Dropout: a simple way to prevent neural networks from overfitting.
\newblock {\em The Journal of Machine Learning Research}, 15(1):1929--1958,
  2014.

\bibitem[\protect\citeauthoryear{Wang \bgroup \em et al.\egroup
  }{2014}]{MMrate}
Senzhang Wang, Xia Hu, Philip~S. Yu, and Zhoujun Li.
\newblock Mmrate: Inferring multi-aspect diffusion networks with multi-pattern
  cascades.
\newblock In {\em Proceedings of the 20th ACM SIGKDD International Conference
  on Knowledge Discovery and Data Mining}, KDD '14, pages 1246--1255, 2014.

\bibitem[\protect\citeauthoryear{Wang \bgroup \em et al.\egroup
  }{2019}]{wang2019deep}
Senzhang Wang, Jiannong Cao, and Philip~S. Yu.
\newblock Deep learning for spatio-temporal data mining: A survey, 2019.

\bibitem[\protect\citeauthoryear{Wooff and Anderson}{2015}]{wooff2015time}
David~A Wooff and Jillian~M Anderson.
\newblock Time-weighted multi-touch attribution and channel relevance in the
  customer journey to online purchase.
\newblock {\em Journal of Statistical Theory and Practice}, 9(2):227--249,
  2015.

\bibitem[\protect\citeauthoryear{Yang \bgroup \em et al.\egroup
  }{2018}]{AAAI18}
Dongdong Yang, Senzhang Wang, and Zhoujun Li.
\newblock Ensemble neural relation extraction with adaptive boosting.
\newblock In {\em Proceedings of the 27th International Joint Conference on
  Artificial Intelligence}, IJCAI'18, pages 4532--4538, 2018.

\bibitem[\protect\citeauthoryear{Zhu \bgroup \em et al.\egroup
  }{2017}]{zhu2017next}
Yu~Zhu, Hao Li, Yikang Liao, Beidou Wang, Ziyu Guan, Haifeng Liu, and Deng Cai.
\newblock What to do next: Modeling user behaviors by time-lstm.
\newblock In {\em Proceedings of the Twenty-Sixth International Joint
  Conference on Artificial Intelligence, IJCAI-17}, pages 3602--3608, 2017.

\end{thebibliography}

\end{document}